\documentclass[aps,prl,showpacs,twocolumn]{revtex4}
\usepackage{amsmath,amsfonts,bm}
\usepackage[dvips]{graphicx}

\begin{document}

\title{
Loop Calculus in Statistical Physics and Information Science}

\author{Michael Chertkov $^{(a)}$ and Vladimir Y. Chernyak $^{(b)}$}

\affiliation{$^{(a)}$ Theoretical Division and Center for Nonlinear
Studies, LANL,  Los Alamos, NM 87545\\
$^{(b)}$ Department of Chemistry, Wayne State University, 5101 Cass Ave, Detroit, MI 48202
}


\date{\today}

\begin{abstract}
Considering a discrete and finite statistical model of a general
position we introduce an exact expression for the partition function
in terms of a finite series. The leading term in the series is the
Bethe-Peierls (Belief Propagation)-BP contribution, the rest are
expressed as loop-contributions on the factor graph and calculated
directly using the BP solution. The series unveils a small parameter
that often makes the BP approximation so successful. Applications of
the loop calculus in statistical physics and information science are
discussed.
\end{abstract}

\pacs{05.50.+q,89.70+C}

\maketitle

Discrete statistical models, the Ising model being the most famous example, play a prominent role
in theoretical and mathematical physics. They are typically defined on a lattice, and major efforts
in the field focused primarily on the case of the infinite lattice size. Similar statistical models
emerge in information science. However, the most interesting questions there are related to graphs
that are very different from a regular lattice. Moreover it is often important to consider large
but finite graphs. Statistical models on graphs with long loops are of particular interest in the
fields of error-correction and combinatorial optimization. These graphs are tree-like locally.

A theoretical approach pioneered by Bethe \cite{35Bet} and Peierls
\cite{36Pei} (see also \cite{82Bax}), who suggested to analyze
statistical models on perfect trees, has largely remained a useful
efficiently solvable toy. Indeed, these models on trees are
effectively one-dimensional, thus exactly-solvable in the
theoretical sense, while computational effort scales linearly with
the generations number. The exact tree results have been extended to
higher-dimensional lattices as uncontrolled approximations. In spite
of the absence of analytical control  the Bethe-Peierls
approximation gives remarkably accurate results, often
out-performing standard mean-field results. The ad-hoc approach was
also re-stated in a variational form \cite{51Kik,05YFW}. Except for
two recent papers \cite{05MR,05PS} that will be discussed later in
the letter, no systematic attempts to construct a regular theory
with a well-defined small parameter and Bethe-Peierls as its leading
approximation have been reported.

A similar tree-based approach in information science has been
developed by Gallager \cite{63Gal} in the context of
error-correction theory. Gallager introduced so called
Low-Density-Parity-Check (LDPC) codes, defined on locally tree-like
Tanner graphs. The problem of ideal decoding, i.e. restoring the
most probable pre-image out of the exponentially large pool of
candidates, is identical to solving a statistical model on the graph
\cite{89Sou}. An approximate yet efficient decoding
Belief-Propagation algorithm introduced by Gallager constitutes an
iterative solution of the Bethe-Peierls equations derived as if the
statistical problem was defined on a tree that locally represents
with the Tanner graph. We utilize this abbreviation coincidence to
call Bethe-Peierls and Belief-Propagation equations by the same
acronym -- BP. Recent resurgence of interest to LDPC codes
\cite{99Mac},  as well as proliferation of the BP approach to other
areas of information and computer science, e.g. artificial
intelligence \cite{88Pea} and combinatorial optimization
\cite{02MPZ}, where interesting statistical models on graphs with
long loops are also involved, posed the following questions. Why
does BP perform so well on graphs with loops? What is the hidden
small parameter that ensures exceptional performance of BP? How can
we systematically correct BP? This letter provides systematic
answers to all these questions.

The letter is organized as follows. We start with introducing notations for a generic statistical
model, formulated in terms of interacting Ising variables with the network described via a factor
graph. We next state our main result: a decomposition of the partition function of the model in a
finite series. The BP expression for the model represents the first term in the series. All other
terms correspond to closed undirected and possibly branching yet not terminating at a node
subgraphs of the factor-graph, referred to as generalized loops. The simplest diagram is a single
loop. An individual contribution is the product of local terms along a generalized loop, expressed
explicitly in terms of simple correlation functions calculated within the BP. We proceed with
discussing the meaning of BP as a successful approximation in terms of the loop series followed by
presenting a clear derivation of the loop series. The derivation includes three steps. We first
introduce a family of local gauge transformations, two per an original Ising variable. The gauge
transformation changes individual terms in the expansion with the full expression for the partition
function natually remaining unchanged. We then fix the gauge in a way that only those terms that
correspondent to generalized loops contribute to the modified series. Finally, we show that the
first term in the resulting generalized loop series corresponds exactly to the standard BP
approximation. This interprets BP as a special gauge choice. We conclude with clarifying the
relation of this work to other recent advances in the subject, and discussing possible applications
and generalizations of the approach.

\begin{figure} [b]
\includegraphics[width=0.35\textwidth]{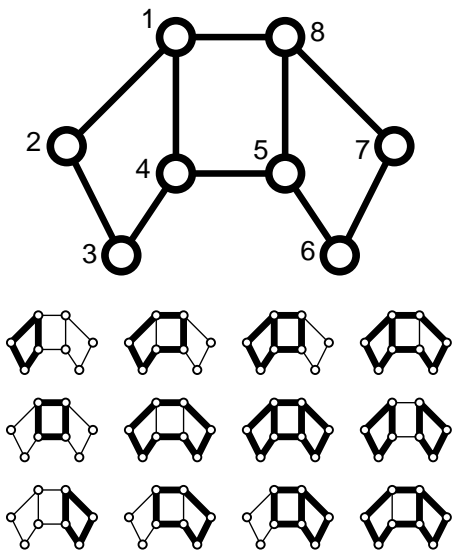}
\caption{Example of a factor graph. Twelve possible marked paths (generalized loops) are shown in
bold in the bottom.} \label{paths}
\end{figure}

{\bf Vertex Model.} Consider a generic discrete statistical model defined for an arbitrary finite
undirected graph, $\Gamma$, with bits $a,b=1,\dots,m$ with the neighbors connected by edges,
$(a,b),\dots$, the neighbor relation expressed as $a\in b$ or $b\in a$. Configurations ${\bm
\sigma}$, are characterized by sets of binary (spin) variables  $\sigma_{ab}=\pm 1$, associated
with the graph edges: ${\bm \sigma}=\{\sigma_{ab};(a,b)\in \Gamma\}$. The probability of
configuration ${\bm\sigma}$ is
\begin{eqnarray}
\label{VM-weight-gen}
 p({\bm\sigma})= Z^{-1}\prod\limits_{a\in \Gamma}f_a({\bm\sigma}_a),\ \
 Z=\sum\limits_{\bm \sigma} \prod\limits_{a\in
 \Gamma}f_a({\bm\sigma}_a),
\end{eqnarray}
$f_a({\bm\sigma}_a)$ being a non-negative function of ${\bm \sigma}_a$ a vector built of
$\sigma_{ab}$ with $b\in a$: ${\bm \sigma}_a=\{\sigma_{ab};b\in a\}$. The notation assumes
$\sigma_{ab}=\sigma_{ba}$. Our vertex model generalizes the celebrated six- and eight-vertex models
of Baxter \cite{82Bax}. An example of a factor graph with $m=8$ that corresponds to
$p(\sigma_1,\sigma_2,\sigma_3,\sigma_4)=Z^{-1}\prod_{a=1}^8f_a({\bm \sigma}_a)$, where
 ${\bm \sigma}_1\equiv\left(\sigma_2,\sigma_4,\sigma_8\right)$,
 ${\bm \sigma}_2\equiv\left(\sigma_1,\sigma_3\right)$,
 ${\bm \sigma}_3\equiv\left(\sigma_2,\sigma_4\right)$,
 ${\bm \sigma}_4\equiv\left(\sigma_1,\sigma_3,\sigma_5\right)$,
 ${\bm \sigma}_5\equiv\left(\sigma_4,\sigma_6,\sigma_8\right)$,
 ${\bm \sigma}_6\equiv\left(\sigma_5,\sigma_7\right)$,
 ${\bm \sigma}_7\equiv\left(\sigma_6,\sigma_8\right)$,
 ${\bm \sigma}_8\equiv\left(\sigma_1,\sigma_5,\sigma_7\right)$,
is shown in Fig.~\ref{paths}.

{\bf Loop decomposition.} The main exact result of the Letter is
decomposition of the partition function defined by
Eq.~(\ref{VM-weight-gen}) in a finite series:
\begin{eqnarray}
 && Z=Z_0\left(1+\sum_{\it C}\frac{\prod\limits_{a \in{\it C}}\mu_a({\it C})}
 {\prod\limits_{(a,b)\in C}(1-m_{ab}({\it C})^2)}\right), \label{Zseries}\\
 && m_{ab}({\it
 C})=\sum_{\sigma_{ab}}\sigma_{ab}b_{ab}(\sigma_{ab}),\label{mab}\\
 && \mu_a=\sum_{{\bm \sigma}_a}\prod_{b\in a,{\it C}}^{b\neq a}(\sigma_{ab}-m_{ab})b_a({\bm
 \sigma}_a),
 \label{mua}
\end{eqnarray}
where summation goes over all allowed (marked) paths ${\it C}$, or
generalized loops. They consist of bits each with at least two
distinct neighbors along the path. Twelve allowed marked paths for
our example are shown in Fig.~(\ref{paths}) on the right. A
generalized loop can be disconnected, e.g. the last one in the
second raw shown in Fig.~(\ref{paths}. In Eqs.~(\ref{Zseries})
$b_{ab}(\sigma_{ab})$, $b_a({\bm \sigma}_a)$ and $Z_0$ are beliefs
(probabilities) defined on edges, bits, and the partition function,
respectively, calculated within the BP. A BP solution can be
interpreted as an exact solution in an infinite tree built by
unwrapping the factor graph. A BP solution can be also interpreted
\cite{05YFW} as a set of beliefs that minimize the Bethe free energy
\begin{equation}
 {\cal F}\!=\!\!\sum_a\!\!\sum_{{\bm \sigma}_a}\!b_a({\bm \sigma}_a)\!\ln\! \frac{b_a({\bm
 \sigma}_a)}{f_a({\bm \sigma}_a)}\!-\!\!\sum_{(a,b)}
 \!\sum_{\sigma_{ab}}\! b_{ab}(\sigma_{ab})\ln b_{ab}(\sigma_{ab}),\nonumber
\end{equation}
under the set of realizability, $0\leq b_a({\bm
\sigma}_a),b_{ab}(\sigma_{ab})\leq 1$, normalization,
$\sum_{{\bf\sigma}_a}b_a({\bm \sigma}_a)=\sum_{\sigma_{ab}}
b_{ab}(\sigma_{ab})=1$, and consistency
$\sum_{{\bm \sigma}_a\backslash\sigma_{ab}}b_a({\bm
 \sigma}_a)=b_{ab}(\sigma_{ab})$, constraints.
The term associated with a marked path is the ratio of the products of irreducable correlation
functions (\ref{mua}) and the quadratic magnetization at-edge functions (\ref{mab}) calculated
along the marked path ${\it C}$ within the BP approximation.

As usual in statistical mechanics exact expressions for the spin
correlation functions can be obtained by differentiating
Eq.~(\ref{Zseries}) with respect to the proper factor functions. In
the tree (no loops) case only the unity term in the r.h.s. of
Eq.~(\ref{Zseries}) survives. In the general case
Eq.~(\ref{Zseries}) provides a clear criterion for the BP
approximation validity: The sum over the loops in the r.h.s. of
Eq.~(\ref{Zseries}) should be small compared to one. The number of
terms in the series increases exponentially with the number of bits.
Therefore, Eq.~(\ref{Zseries}) becomes useful for selecting a
smaller than exponential number of leading contributions. In a large
system the leading contribution comes from the paths with the number
of degree two connectivity nodes substantially exceeding the number
of branching nodes, i.e. the ones with higher connectivity degree.
According to Eq.~(\ref{Zseries}) the contribution of a long path is
given by the ratio of the along-the-path product of the irreducible
nearest-neighbor spin correlation functions associated with a bit,
$\mu_a$ to the along-the-path product of the edge contributions,
$1/(1-m_{ab}^2)$. All are calculated within BP. Therefore, the small
parameter in the perturbation theory is $\varepsilon=\prod_{a\in
{\it C}}\mu_a({\it C})/\prod_{(a,b)\in {\it C}} (1-m_{ab}^2)$. If
$\varepsilon$ is much smaller than one for all marked paths the BP
approximation is valid. We anticipate the loop formula
(\ref{Zseries}) to be extremely useful for analysis and possible
differentiation between the loop contributions. Whether the series
is dominated by a single loop contribution or some number of
comparable loop correction, will depend on the problem specifics
(form of the factor graph and functions). In the former case the
leading correction to the BP result is given by the marked path with
the largest $\varepsilon$.

{\bf Derivation of the loop formula.} We relax the condition $\sigma_{ab}=\sigma_{ba}$ in
Eq.~(\ref{VM-weight-gen}) and treat $\sigma_{ab}$ and $\sigma_{ba}$ as independent variables. This
allows to represent the partition function in the form
\begin{eqnarray}
 \label{Z-gen-graph} Z=\sum_{{\bm \sigma}'}
 \prod_{a}f_{a}({\bm\sigma}_{a})\prod_{(b,c)}\frac{1+\sigma_{bc}\sigma_{cb}}{2},
\end{eqnarray}
where there are twice more components since any pair of variables $\sigma_{ab}$ and $\sigma_{ba}$
enters ${\bm \sigma}$ independently. It is also assumed in Eq.~(\ref{Z-gen-graph}) that each edge
contributes to the product over $(b,c)$ only once. The representation (\ref{Z-gen-graph}) is
advantageous over the original one (\ref{VM-weight-gen}) since ${\bm\sigma}_{a}$ at different bits
become independent.  We further introduce a parameter vector ${\bm\eta}$ with independent
components $\eta_{ab}$ (i.e., $\eta_{ab}\neq\eta_{ba}$). Making use of the key identity
\begin{eqnarray} &&
\frac{\cosh(\eta_{bc}+\eta_{cb})(1+\sigma_{bc}\sigma_{cb})}{(\cosh\eta_{bc}+\sigma_{bc}\sinh\eta_{bc})
(\cosh\eta_{cb}+\sigma_{cb}\sinh\eta_{cb})} =V_{bc},\nonumber\\ &&
V_{bc}\left(\sigma_{bc},\sigma_{cb}\right)\!=\!1\!+\!
\left(\sinh(\eta_{bc}+\eta_{cb})\!-\!\sigma_{bc}\cosh(\eta_{bc}\!+\!\eta_{cb})\right)\nonumber\\
&&\times
\left(\sinh(\eta_{bc}\!+\!\eta_{cb})\!-\!\sigma_{cb}\cosh(\eta_{bc}\!+\!\eta_{cb})\right),
\label{Vbc}
\end{eqnarray}
we transform the product over edges on the rhs of Eq.~(\ref{Z-gen-graph}) to arrive at:
\begin{eqnarray}
\label{Z-ready-gen}
 &&
 Z\!=\!\left(\prod_{(b,c)}2\cosh\left(\eta_{bc}\!+\!\eta_{cb}\right)\right)^{-1}\!
 \sum_{\bm\sigma'}\prod_{a}P_{a}\prod_{bc}V_{bc},\\
 &&
  P_{a}({\bm\sigma}_{a})=f_{a}({\bm\sigma}_{a})\prod_{b\in a}
 \left(\cosh\eta_{ab}+\sigma_{ba}\sinh\eta_{ab}\right).
\end{eqnarray}
The desired decomposition Eq.~(\ref{Zseries}) is obtained by choosing some special values for the
$\eta$-variables (fixing the gauge !!) and expanding the $V$-terms in Eq.~(\ref{Z-ready-gen}) in a
series followed by a local computation (summations over $\sigma$-variables at the edges).
Individual contributions to the series are naturally identified with subgraphs of the original
graph defined by a simple rule: Edge $(a,b)$ belongs to the subgraph if the corresponding ``vertex"
$V_{ab}$ on the rhs of Eq.~(\ref{Z-ready-gen}) contributes using its second (non-unity) term,
naturally defined according to Eq.~(\ref{Vbc}). We next utilize the freedom in the choice of
${\bm\eta}$. The contributions that originate from subgraphs with loose ends vanish provided the
following system of equations is satisfied:
\begin{eqnarray}
\label{General-BP-gen}
\sum_{{\bm\sigma}_{a}}\left(\tanh(\eta_{ab}+\eta_{ba})-\sigma_{ba}\right)P_{a}({\bm\sigma}_{a})=0.
\end{eqnarray}
The number of equations is exactly equal to the number of $\eta$ variables. Moreover,
Eqs.~(\ref{General-BP-gen}) are nothing but BP equations: simple algebraic manipulations (see
\cite{06CCb} for details) allow to recast Eq.~(\ref{General-BP-gen}) in a more traditional BP form
\begin{eqnarray}
 \nonumber
 \tanh\eta_{ba}\!=\!
 \frac{\sum_{{\bm\sigma}_{a}}\sigma_{ab}f_{a}({\bm\sigma}_{a})\prod_{c\in
 a}^{c\ne
 b}\left(\cosh\eta_{ac}+\sigma_{ac}\sinh\eta_{ac}\right)}{\sum_{{\bm\sigma}_{a}}
 f_{a}({\bm\sigma}_{a})\prod_{c\in a}^{c\ne
 b}\left(\cosh\eta_{ac}+\sigma_{ac}\sinh\eta_{ac}\right)},
\end{eqnarray}
with the relation between the beliefs that minimize the Bethe free energy ${\cal F}$ and the $\eta$
fields according to:
\begin{eqnarray}
b_a({\bm \sigma}_a)=\frac{P_a({\bm\sigma}_a)}{\sum_{{\bm
\sigma}_a}P_a({\bm\sigma}_a)}.\nonumber
\end{eqnarray}
The final  expression Eq.~(\ref{Zseries}) emerges as a result of direct expansion of the $V$ term
in Eq.~(\ref{Z-gen-graph}), performing summations over local $\sigma$-variables, making use of
Eqs.~(\ref{mab},\ref{mua}), and also identifying the BP expression for the partition function as
\begin{eqnarray}
Z_0=\frac{\prod_{a}P_{a}({\bm\sigma}_a)
}{\prod_{(b,c)}2\cosh\left(\eta_{bc}+\eta_{cb}\right)}.\nonumber
\end{eqnarray}
To summarize, Eq.~(\ref{Zseries}) represents a finite series where all individual contributions are
related to the corresponding generalized loops. This fine feature is achieved via a special
selection of the BP gauge (\ref{General-BP-gen}). The condition enforces the ``no loose ends" rule
thus prohibiting anything but generalized loop contributions to Eq.~(\ref{Zseries}). Any individual
contribution is expressed explicitly in terms of the BP solution.

{\bf Comments, Conclusions and Path Forward.} We expect that BP equations may have multiple
solutions for the model with loops. This expectation naturally follows from the notion of the
infinite covering graph, as different BP solutions correspond to different ways to spontaneously
break symmetry on the infinite structure. This different BP solutions will generate loop series
(\ref{Zseries}) that are different term by term but give the same result for the sum. Finding the
``optimal" BP solution with the smallest $\varepsilon$, characterizing loop correction to the BP
solution, is important for applications. A solution related to the absolute minimum of the Bethe
free energy would be a natural candidate. However, one cannot guarantee that the absolute minimum,
as opposed to other local minima of ${\cal F}$ is always ``optimal" for arbitrary $f_\alpha$.

We further briefly discuss other models related to the general one
discussed in the paper.  The vertex model can be considered  on a
graph of the special oriented/biparitite type. A bipartite graph
contains two families of nodes, referred to as bits and checks, so
that the neighbor relations occur only between the nodes from
opposite families. A bipartite factor-graph model with an additional
property that any factor associated with a bit is nonzero only if
all Ising variables at the neighboring edges are the same, leads to
the factor-graph model considered in \cite{05YFW}. Actually, this
factorization condition means re-assignment of the Ising variables,
defined at the edges of the original vertex model, to the
corresponding bits of the bipartite factor-graph model. Furthermore,
if only checks of degree two (each connected to only two bits) are
considered, the bipartite factor graph model is reduced to the
standard binary-interaction Ising model. The loop series derived in
this Letter is obviously valid for all less general aforementioned
models. Also note that the bipartite factor graph model was chosen
in \cite{06CCb} to introduce an alternative derivation of the loop
series via an integral representation, where BP corresponds to the
saddle-point approximation for the resulting integral.

Let us now comment on two relevant papers \cite{05MR,05PS}. The
Ising model on a graph with loops has been considered by Montanari
and Rizzo \cite{05MR}, where a set of exact equations has been
derived that relates the correlation functions to each other. This
system of equations is under-defined, however, if irreducible
correlations are neglected the BP result is restored. This feature
has been used \cite{05MR} to generate a perturbative expansion for
corrections to BP in terms of irreducible correlations. A
complementary approach for the Ising model on a lattice has been
taken by Parisi and Slanina \cite{05PS}, who utilized an integral
representation developed by Efetov \cite{90Efe}. The saddle-point
for the integral representation used in \cite{05PS} turns out to be
exactly the BP solution. Calculating perturbative corrections to
magnetization, the authors of \cite{05PS} encountered divergences in
their representation for the partition function, however, the
divergences cancelled out from the leading order correction to the
magnetization revealing a sensible loop correction to BP. These
papers, \cite{05MR} and \cite{05PS}, became important initial steps
towards calculating and understanding loop corrections to BP.
However, both approaches are very far from being complete and
problems-free. Thus, \cite{05MR} lacks an invariant representation
in terms of the partition function, and requires operating with
correlation functions instead. Besides, the complexity of the
equations related to the higher-order corrections rapidly grows with
the order. The complimentary approach of \cite{05PS} contains
dangerous, since lacking analytical control, divergences (zero
modes), which constitutes a very problematic symptom for any field
theory. Both \cite{05MR} and \cite{05PS} focus on the Ising
pair-wise interaction model. The extensions of the proposed methods
to the most interesting from the information theory viewpoint
multi-bit interaction cases do not look straightforward. Finally,
the approaches of \cite{05MR} and \cite{05PS}, if extended to
higher-order corrections, will result in infinite series. Re-summing
the corrections in all orders, so that the result is presented in
terms of a finite series, does not look feasible within the proposed
techniques.

We conclude with a discussion of possible applications and generalizations. We see a major utility
for Eq.~(\ref{Zseries}) in its direct application to the models without short loops. In this case
Eq.~(\ref{Zseries}) constitutes an efficient tool for improving BP through accounting for the
shortest loop corrections first and then moving gradually (up to the point when complexity is still
feasible) to account for longer and longer loops. Another application of Eq.~(\ref{Zseries}) is
direct use of $\varepsilon$ as a test parameter for the BP approximation validity: If the shortest
loop corrections to BP are not small one should either look for another solution of BP (hoping that
the loop correction will be small within the corresponding loop series) or conclude that no
feasible BP solution, resulting in a small $\varepsilon$, can be used as a valid approximation.
There is also a strong generalization potential here. If a problem is multi-scale with both short
and long loops present in the factor graph, a development of a synthetic approach combining
Generalized Belief Propagation approach of \cite{05YFW} (that is efficient in accounting for local
correlations) and a corresponding version of Eq.~(\ref{Zseries}) can be beneficial. Finally, our
approach can be also useful for analysis of standard (for statistical physics and field theory)
lattice problems. A particularly interesting direction will be to use Eq.~(\ref{Zseries}) for
introducing a new form of resummation of different scales. This can be applied for analysis of the
lattice models at the critical point where correlations are long-range.

We are thankful to M. Stepanov for many fruitful discussions. The
work at LANL was supported by LDRD program, and through start-up
funds at WSU.

\end{document}